\documentstyle[12pt,epsfig]{article}
\let\section=\subsection     \let\subsection=\subsubsection                %%
\newcommand{\beq}{\begin{equation}}
\newcommand{\eeq}{\end{equation}}
\newcommand{\bea}{\begin{eqnarray}}
\newcommand{\eea}{\end{eqnarray}}
\newcommand{\bce}{\begin{center}}
\newcommand{\ece}{\end{center}}
\newcommand{\eg}{{\it e.g.}}
\newcommand{\ie}{{\it i.e.}}
\newcommand{\etal}{{\it et al.}}
\begin{document}
\begin{center}
   {\large \bf INTERMEDIATE-MASS DILEPTONS}\\[2mm]
   {\large \bf AT THE CERN-SPS AND RHIC}\\[5mm]
   RALF RAPP \\[5mm]
   {\small \it  Department of Physics and Astronomy \\
   SUNY Stony Brook, New York 11794-3800, USA \\[8mm] }
\end{center}

\begin{abstract}\noindent
 The significance of thermal dilepton radiation at intermediate
 invariant masses (1~GeV~$<M<$~3~GeV) in ultrarelativistic 
 heavy-ion collisions is investigated. At CERN-SpS energies, a 
 consistent explanation of the excess observed by NA50 can be given.  
 At RHIC energies the thermal signal is dominated by early emission
 indicative for QGP formation. Chemical under-saturation effects 
 and the competition with open-charm contributions are addressed. 
\end{abstract}

%%%%%%%%%%%%%%%%%%%%%%
\section{Introduction}
%%%%%%%%%%%%%%%%%%%%%%
Due to their penetrating nature dilepton probes are among the 
most promising observables to access the high temperature/density zones
formed in the early phases of (ultra-) relativistic heavy-ion collisions
(URHIC's).
In the low-mass region (LMR, $M\le$~1~GeV) dilepton emission
is governed by the light vector mesons $\rho$, $\omega$ and $\phi$
attaching the main interest to their medium modifications and possible
signatures for the restoration of chiral symmetry in strongly
interacting matter. In the high-mass region (HMR, $M\ge$~3~GeV), the focus
is on the dissolution of the heavy quarkonium bound states ($J/\Psi$, 
$\Upsilon$) to detect the onset of deconfinement. In this talk
we will address the intermediate-mass region (IMR) between the $\phi$
and $J/\Psi$ (1~GeV~$\le M \le$~3~GeV). Here, the thermal dilepton production
rate is essentially structureless and can be rather well approximated 
by perturbative $q\bar q \to l^+l^-$ annihilation 
for both hadronic and quark-gluon phases (as can be inferred 
from the well-known total cross section $\sigma(e^+e^-\to hadrons)$ above
$M\simeq$~1.5~GeV). 
With final-state hadron decays being concentrated in the LMR, the main
competitors with thermal radiation from an interacting fireball 
are primordial processes, most notably Drell-Yan (DY) annihilation and 
the simultaneous decays of associatedly produced open-charm 
(or bottom) mesons, \eg,  
$D\to l^-\bar\nu X$ and $\bar D\to l^+\nu X$.   
An excess over these from proton-proton collision extrapolated sources 
has long been proposed as a suitable signal   
for the early high-temperature phases in URHIC's~\cite{Shu80}. 
Since the expected temperatures $T\ll M$ in the IMR, the  
thermal signal might be sufficiently sensitive to reflect the initial 
temperature and lifetime of a possibly formed 
Quark-Gluon Plasma (QGP).  

In the following we will investigate these issues in the context
of data from the  CERN-SpS (Sect.~\ref{sec_sps}) and then  
apply our current understanding to assess upcoming measurements at RHIC 
(Sect.~\ref{sec_rhic}). We finish with some concluding
remarks in Sect.~\ref{sec_concl}.

%%%%%%%%%%%%%%%%%%%%%%%%%%%%%%%%%%%
\section{I-M Dileptons at the SpS}
\label{sec_sps}
%%%%%%%%%%%%%%%%%%%%%%%%%%%%%%%%%%%

%%%%%%%%%%%%%%%%%%%%%%%%%%%%%%%%%%%%%%%%%%%%%%%%%%%%%%%%%%%%%%%%%%%
\subsection{Experimental Results and Previous Theoretical Analyses}
%%%%%%%%%%%%%%%%%%%%%%%%%%%%%%%%%%%%%%%%%%%%%%%%%%%%%%%%%%%%%%%%%%%
At the CERN-SpS I-M dilepton spectra have been measured by the 
NA38/50 and HELIOS-3 collaborations. Both have found a significant
excess of a factor $\sim$~2 in central $A$-$A$ collisions over 
open-charm and Drell-Yan sources scaled up from $p$-$A$ systems. 

\begin{figure}[!htb]
\epsfig{file=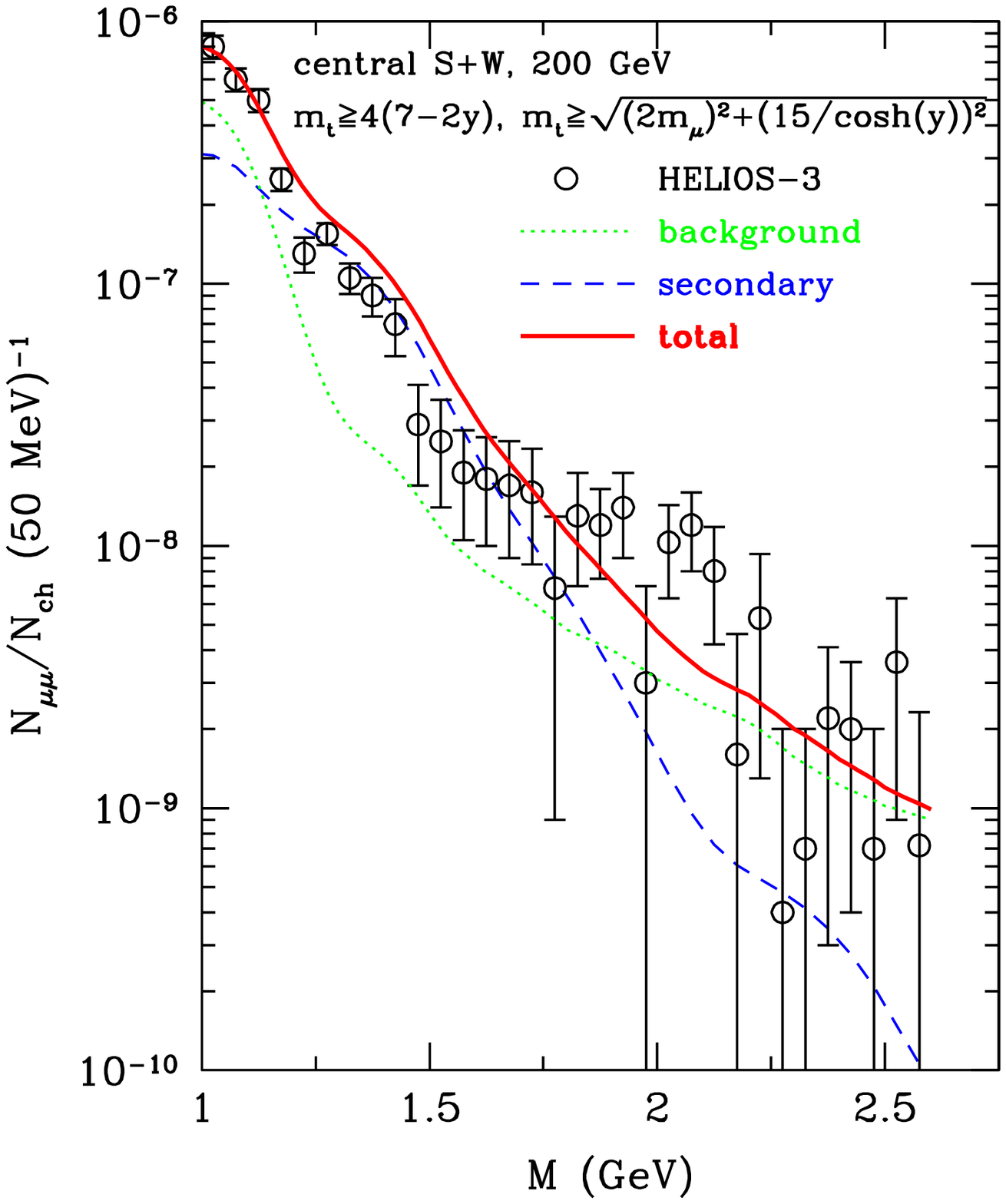,width=7.4cm}
\hspace{-1.1cm}
\epsfig{file=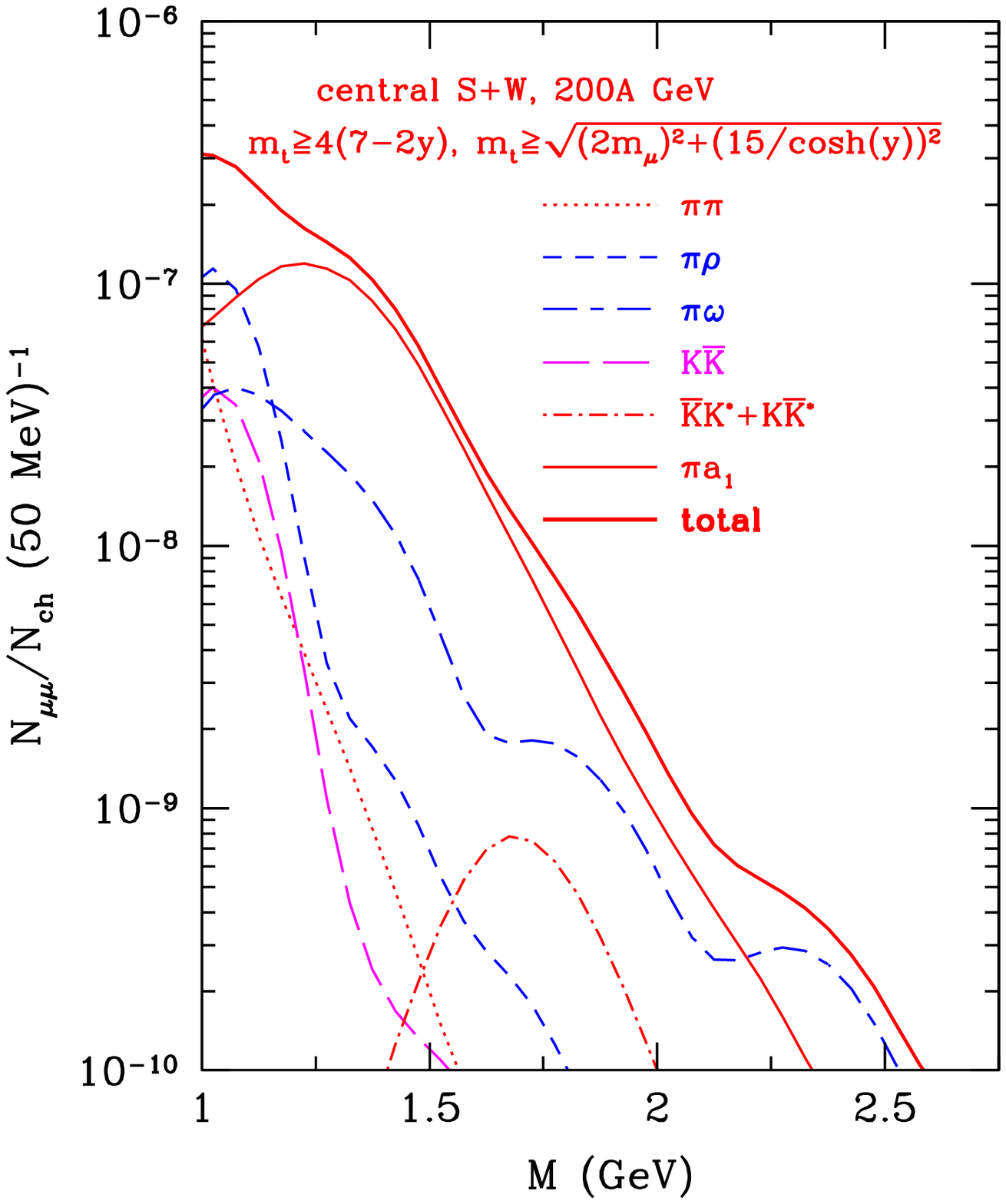,width=7.4cm}
\vspace{-1cm}
\caption{\small{Left panel: HELIOS-3 dimuon data~\protect\cite{HELIOS3}
from central S+W compared to the 'standard' background (consisting of 
Drell-Yan and open charm) and the additional yield from secondary 
hadronic annihilation processes evaluated within  a transport 
model~\protect\cite{LG98}; right panel: decomposition of the secondary 
reactions~\protect\cite{LG98}.}}
\label{fig_helios3}
\end{figure}

Fig.~\ref{fig_helios3} shows a comparison of the HELIOS-3 data
with transport calculations of Li and Gale~\cite{LG98}: apparently
the additional yield from in-medium hadronic annihilation processes
satisfactorily explains the data (left panel); 
note that the prevailing 
channels, $\pi a_1$ and $\pi\omega$, are of '4-pion type' (right panel).
Within the hadronic transport framework, QGP formation could not be
explicitly addressed.
In the context of the NA50 data~\cite{Scomp98,Bord99} 
other possibilities
for the origin of the additional yield have been elaborated: 
\begin{itemize}
\item[(i)]
the NA50 collaboration pointed out that an enhancement of the open-charm
contribution by a factor of $\sim$~3 gives a good account of the data
in central Pb(158~AGeV)+Pb. From the theoretical side, however,
such an effect is difficult to justify; 
\item[(ii)]
Lin and Wang investigated 
whether a broadening of transverse-momen\-tum distributions 
due to a (strong) rescattering of charm quarks in matter might
enrich the yield within the NA50 acceptance~\cite{LW98}. The resulting
increase amounts to about 20\%~\cite{Bord99};
\item[(iii)] 
Spieles \etal~\cite{Spiel98} evaluated 'secondary' 
Drell-Yan processes (\eg, $\pi N\to l^+l^-X$) arising in the later 
stages of the collision and found 
a 10\% enhancement of the primordial Drell-Yan around $M\simeq$~2~GeV; 
\item[(iv)]
thermal radiation from an equilibrated expanding 
fireball~\cite{GKP99,RS99}. 
\end{itemize}
In the following, we will pursue the last item in more detail.  

%%%%%%%%%%%%%%%%%%%%%%%%%%%%%%%%%%%%%%%%%%%%%%%%%%%%%%%
\subsection{Drell-Yan Annihilation and NA50 Acceptance}
\begin{figure}
\begin{minipage}[t]{6.7cm}
\vspace{-0.4cm}
\bce
\hspace{-1cm}
\epsfig{file=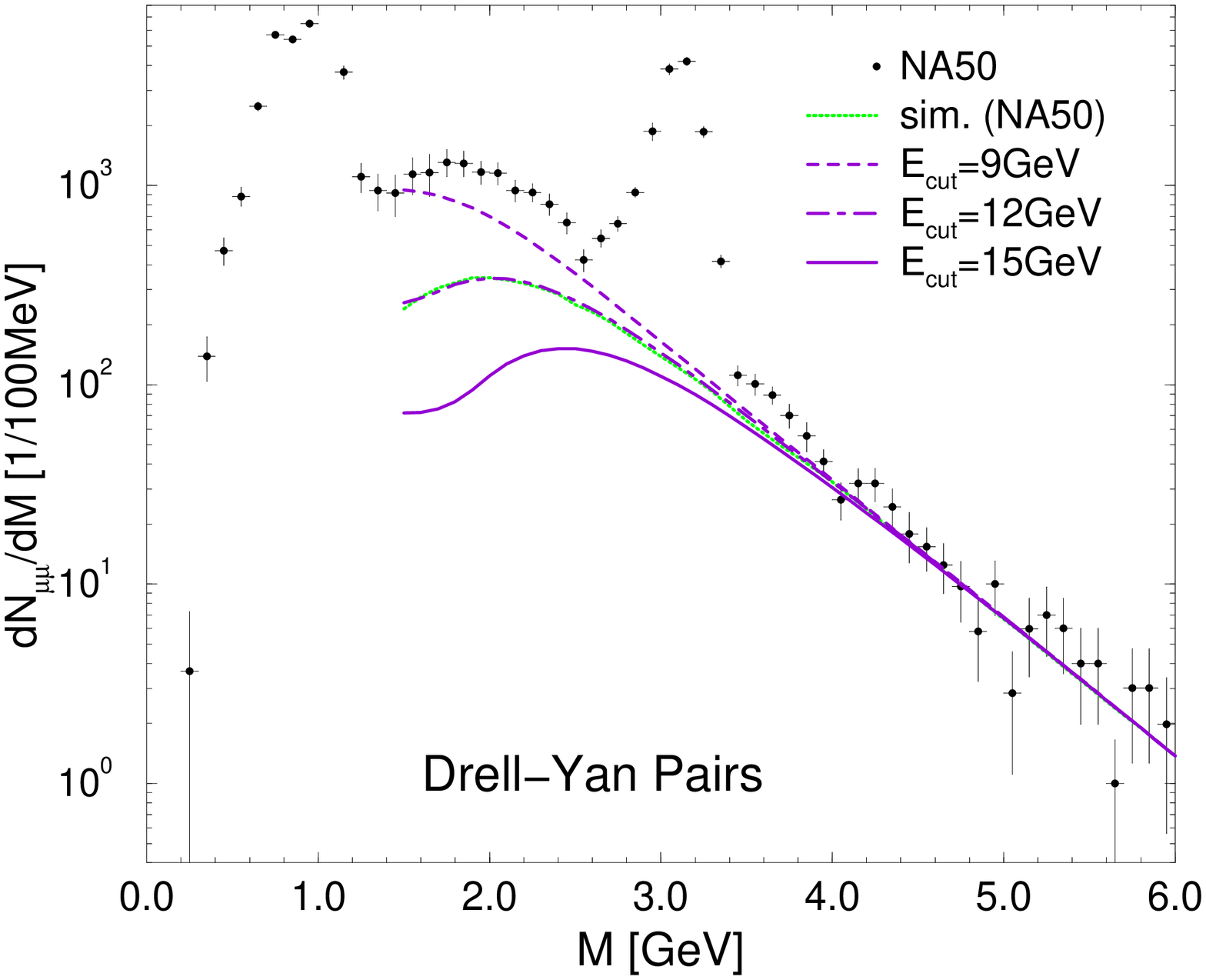,width=7.2cm,angle=-90}
\ece
\vspace{-0.45cm}
\caption{{\small DY-calculations
compared to NA50 simulations (dotted line) and 
data~\protect\cite{Scomp98,Bord99} in central Pb(158AGeV)+Pb.}}
\label{fig_dycuts}
\end{minipage}
\hspace{0.2cm}
\begin{minipage}[t]{6.7cm}
\bce
\epsfig{file=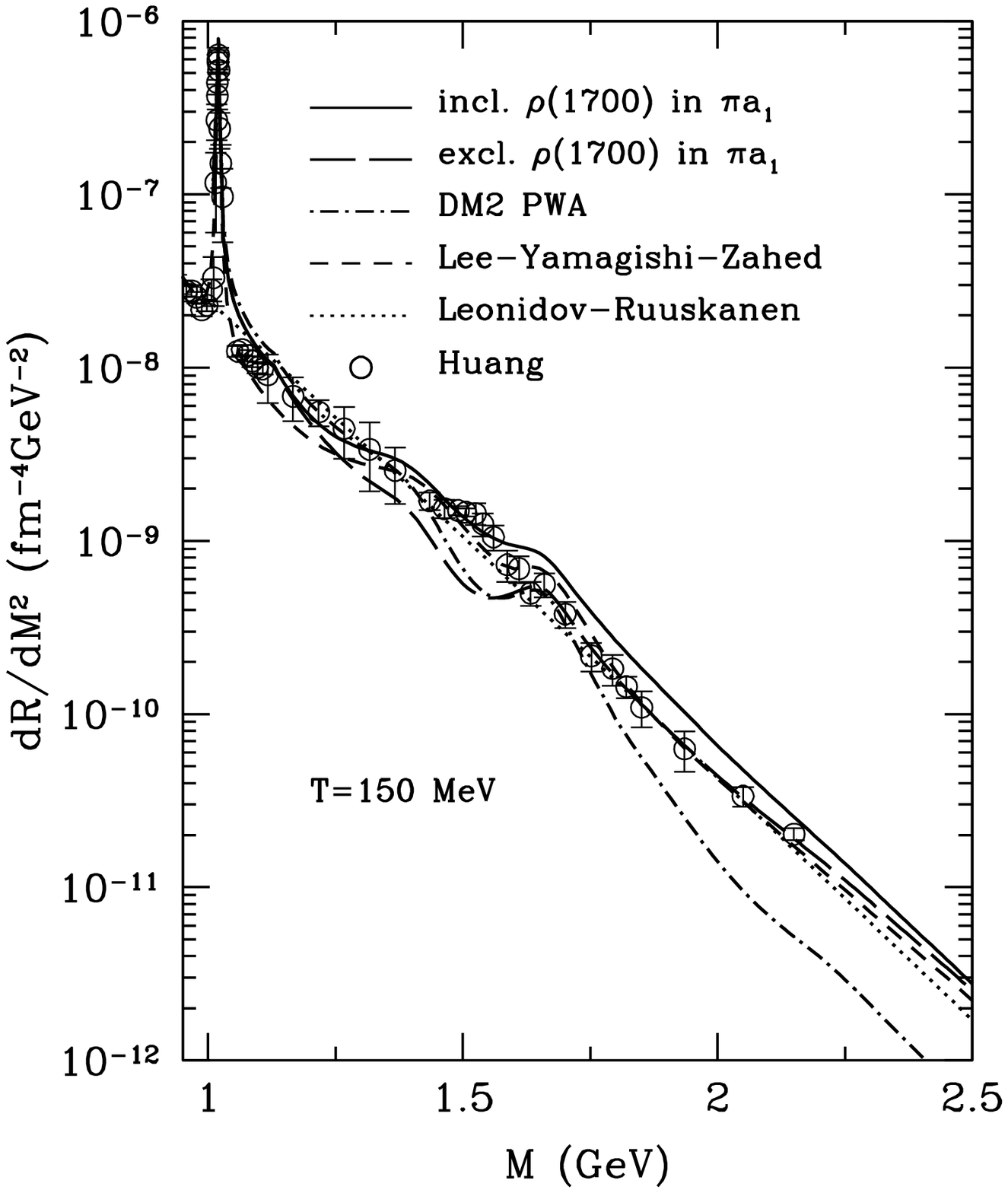,width=7cm,height=6.1cm}
\ece
\vspace{-1cm}
\caption{{\small Compilation~\protect\cite{LG98} of various approaches
to assess the thermal dilepton production rate in the IMR.}}
\label{fig_imrates}
\end{minipage}
\end{figure}

%%%%%%%%%%%%%%%%%%%%%%%%%%%%%%%%%%%%%%%%%%%%%%%%%%%%%%%
To enable a comparison of our final results with  
NA50 data we need to determine their normalization and
acceptance corrections. To this end we use the primordial
Drell-Yan contribution to achieve this. For central 
$A$-$A$ collisions it is given by
\beq
\frac{dN_{DY}^{AA}}{dM dy}(b=0)=\frac{3}{4\pi R_0^2} \ A^{4/3} \ 
\frac{d\sigma_{DY}^{NN}}{dM dy} \ , 
\label{dyaa}
\eeq
which we also employ for slightly non-central ones with
an accordingly reduced $A=N_{part}/2$. Exploiting the fact that the 
high-mass tail ($M$~$\ge$~4~GeV) of the data is entirely saturated 
by DY-pairs we obtain the overall normalization which will
also be applied to the thermal production. We furthermore use 
the calculated DY-spectrum to test our approximate 
acceptance: in addition to geometric 
cuts~\cite{Scomp99} on the single-muon tracks imposed by the NA50
spectrometer set-up, the muons experience substantial absorption 
when traversing the hadron absorber. The latter can be roughly
represented by a lower energy cutoff which is determined~\cite{RS99} 
by requiring to reproduce the DY-results of NA50 detector simulations, 
cf. Fig.~\ref{fig_dycuts}.

%%%%%%%%%%%%%%%%%%%%%%%%%%%%%%%%%%%%%%%%%%%%%%%%%%%%%%%%%%%%%
\subsection{Thermal Rates, Space-Time Evolution and Spectra}
%%%%%%%%%%%%%%%%%%%%%%%%%%%%%%%%%%%%%%%%%%%%%%%%%%%%%%%%%%%%%
Based on the assumption that an interacting fireball formed
in heavy-ion collisions is in local thermal equilibrium 
(in the 'comoving' frame of expansion), the evaluation
of the thermal dilepton yield requires two ingredients:
production rates and the time evolution of volume/temperature. 
In the IMR the former turns out to be given 
in a rather model-independent way by the result from
perturbative QCD for the $q\bar q$ annihilation process,
\beq
\frac{d^8N_{\mu\mu}^{therm}}{d^4xd^4q} = \frac{\alpha^2}{4\pi^4} f^B(q_0;T)
\sum\limits_{q=u,d,s} (e_q)^2 \ + \ {\cal O}(\alpha_s) 
\eeq
for both QGP and hadron gas (HG) phases. This is a direct consequence 
of the well-known 'duality' threshold located around 1.5~GeV in the 
inverse process of $e^+e^-\to hadrons$ annihilation.  
It is further corroborated by explicit hadronic rate calculations
as evidenced from the compilation displayed in 
Fig.~\ref{fig_imrates}~\cite{LG98}. 
$\alpha_s$-corrections may be as large as 
20-30\%, whereas higher order temperature/density  
effects are smaller being of order ${\cal O}(T/M)$, ${\cal O}(\mu_q/M)$.

\begin{figure}[!tb]
\vspace{-1.2cm}
\bce
\epsfig{file=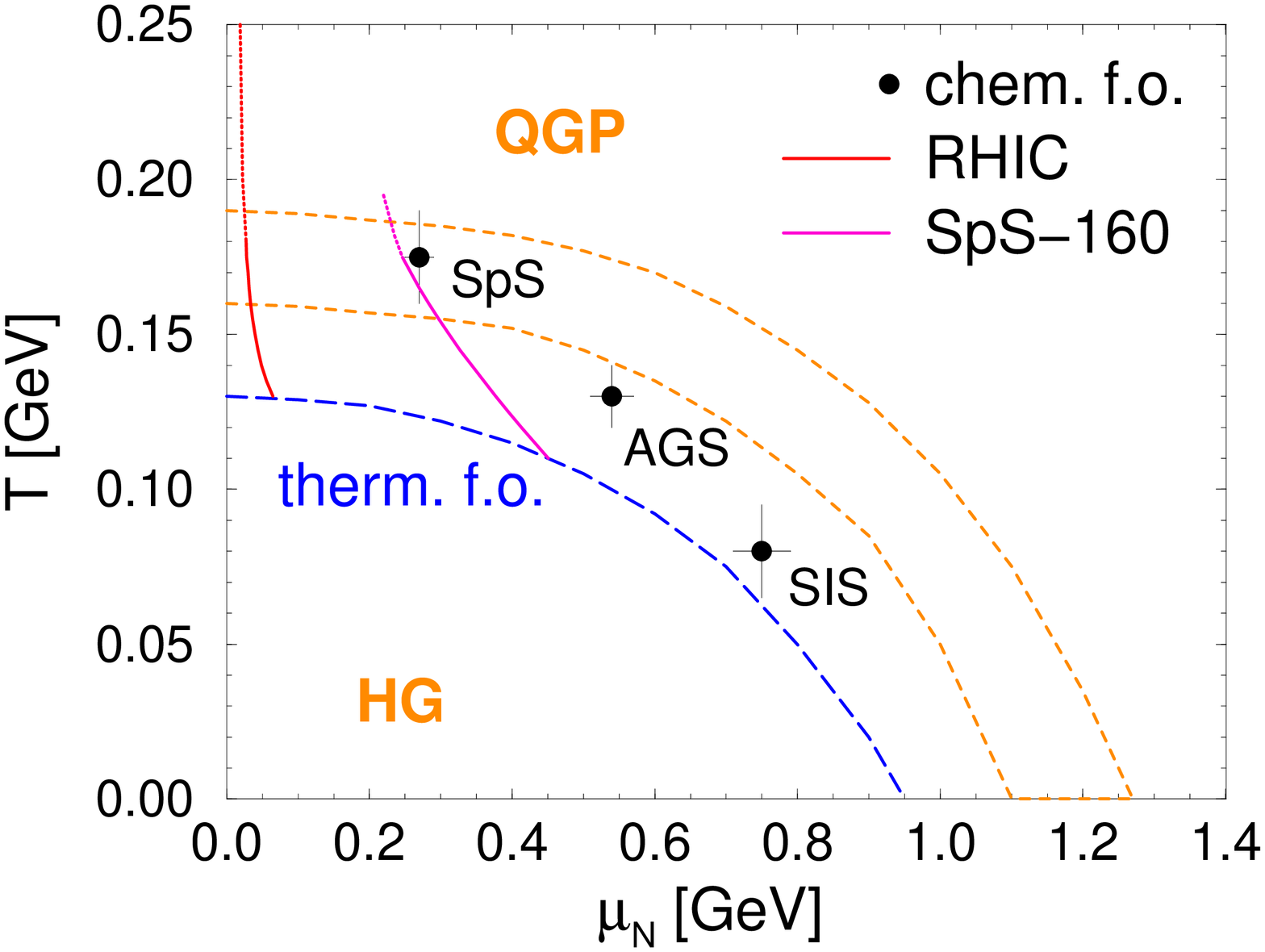,width=6.5cm,angle=-90}
\epsfig{file=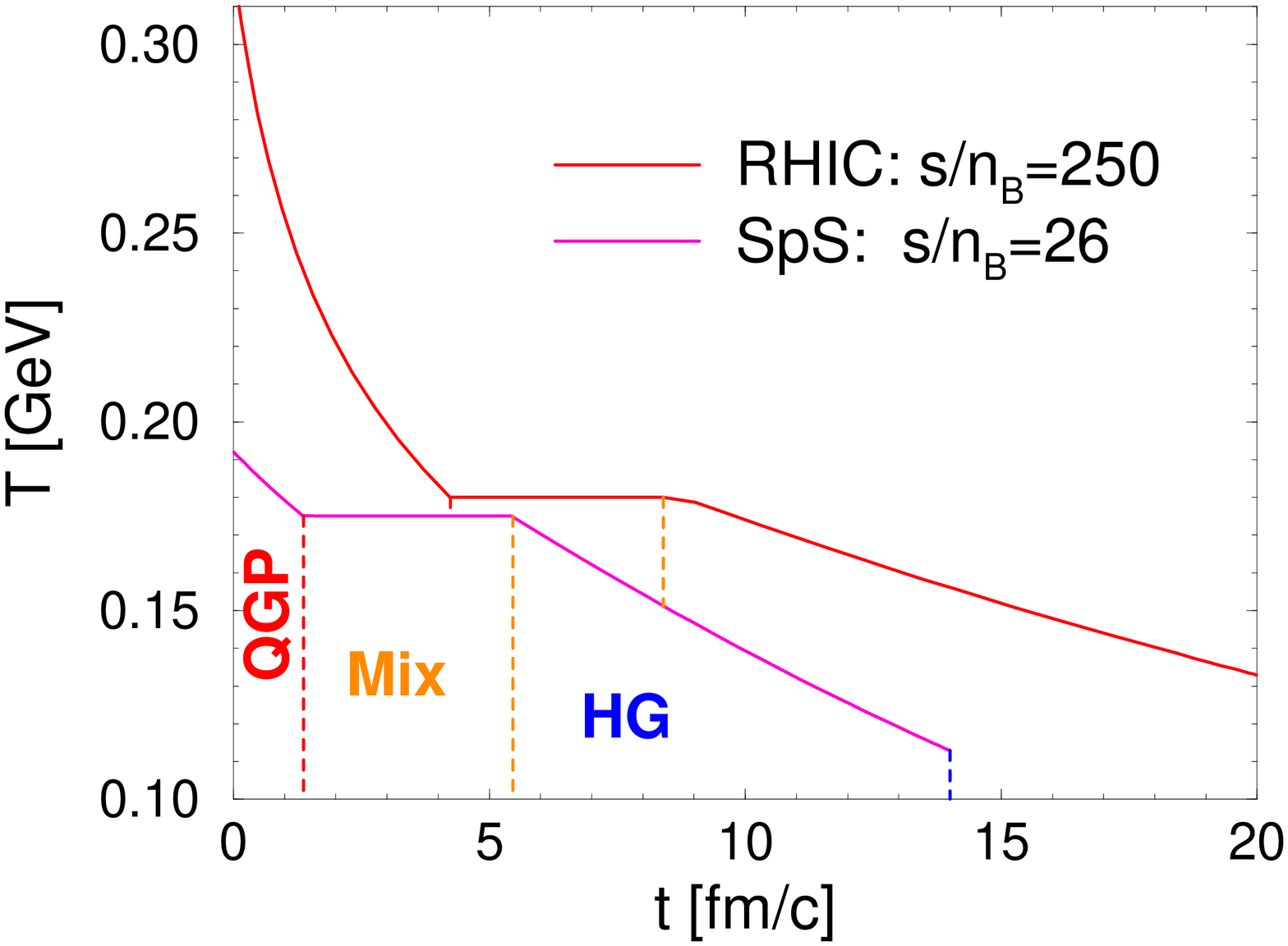,width=6.5cm,angle=-90}
\ece
\vspace{-0.7cm}
\caption{{\small Space-time description of central heavy-ion 
reactions at SpS and RHIC energies within an expanding
thermal fireball model.}}
\label{fig_timeevo}
\end{figure}

For the space-time evolution of central $A$-$A$ collisions we employ
an expanding thermal fireball model~\cite{RW99,RW00} which is based 
on an ideal-QGP and resonance-hadron-gas equation of state. 
Entropy and baryon-number conservation fix a trajectory 
in the $T$-$\mu_N$ plane with the ratio $s/n_B$ chosen in accord with 
experimental
information on chemical freezeout at the SpS~\cite{pbm98}, cf.~left panel
of Fig.~\ref{fig_timeevo}. 
Pion- and kaon-number conservation ensure the correct particle
abundances at thermal freezeout towards which finite chemical
potentials $\mu_\pi$, $\mu_{K,\bar K}$ build up.
A time scale is introduced through a hydro-type volume expansion
which yields realistic final flow velocities and transverse
sizes~\cite{RW99,RS99}.
A QGP-HG mixed phase is constructed from standard
entropy balancing resulting in a temperature evolution shown
in the right panel of Fig.~\ref{fig_timeevo}.

\begin{figure}[!b]
\vspace{-1.2cm}
\bce
\epsfig{file=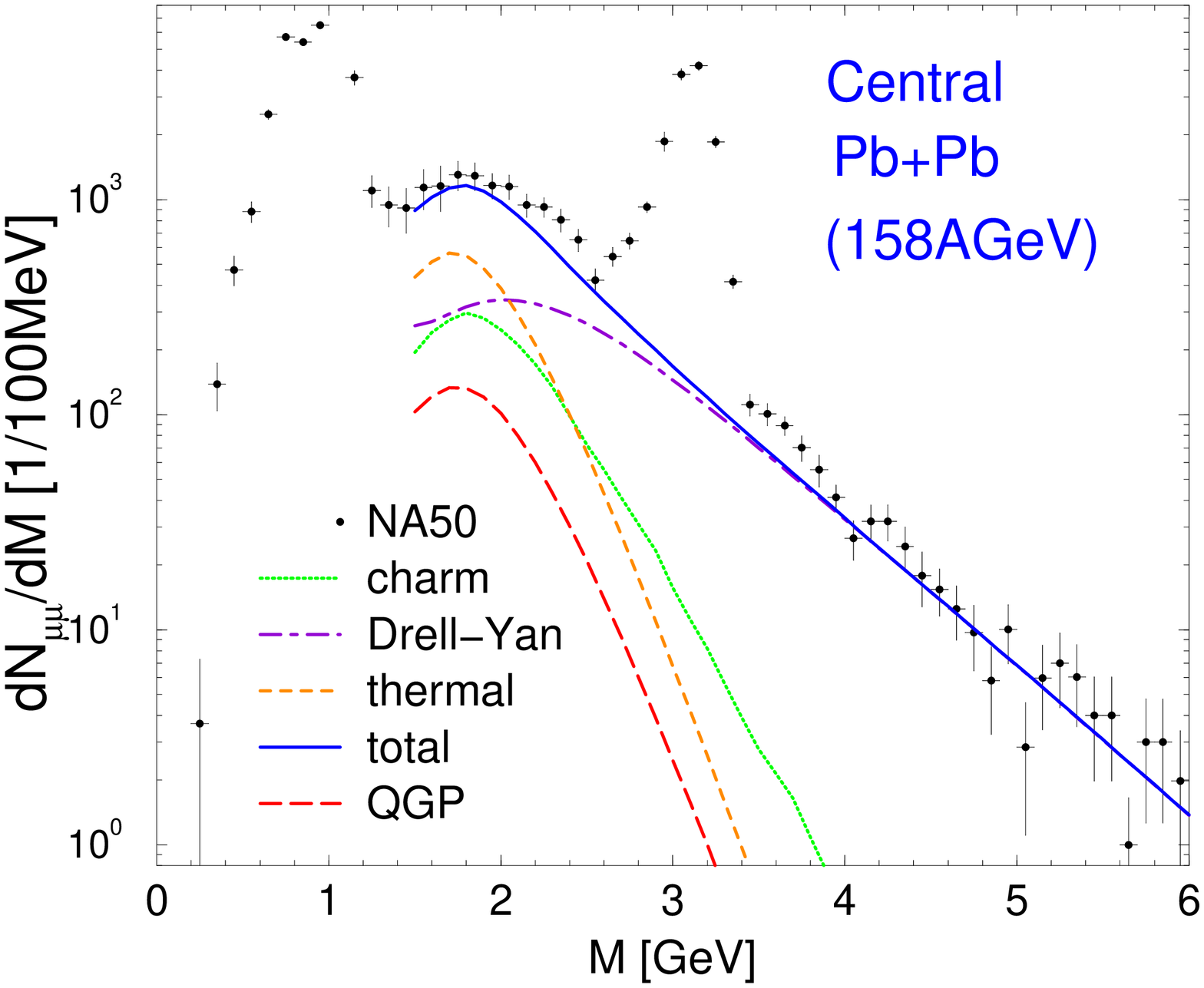,width=6.5cm,angle=-90}
\epsfig{file=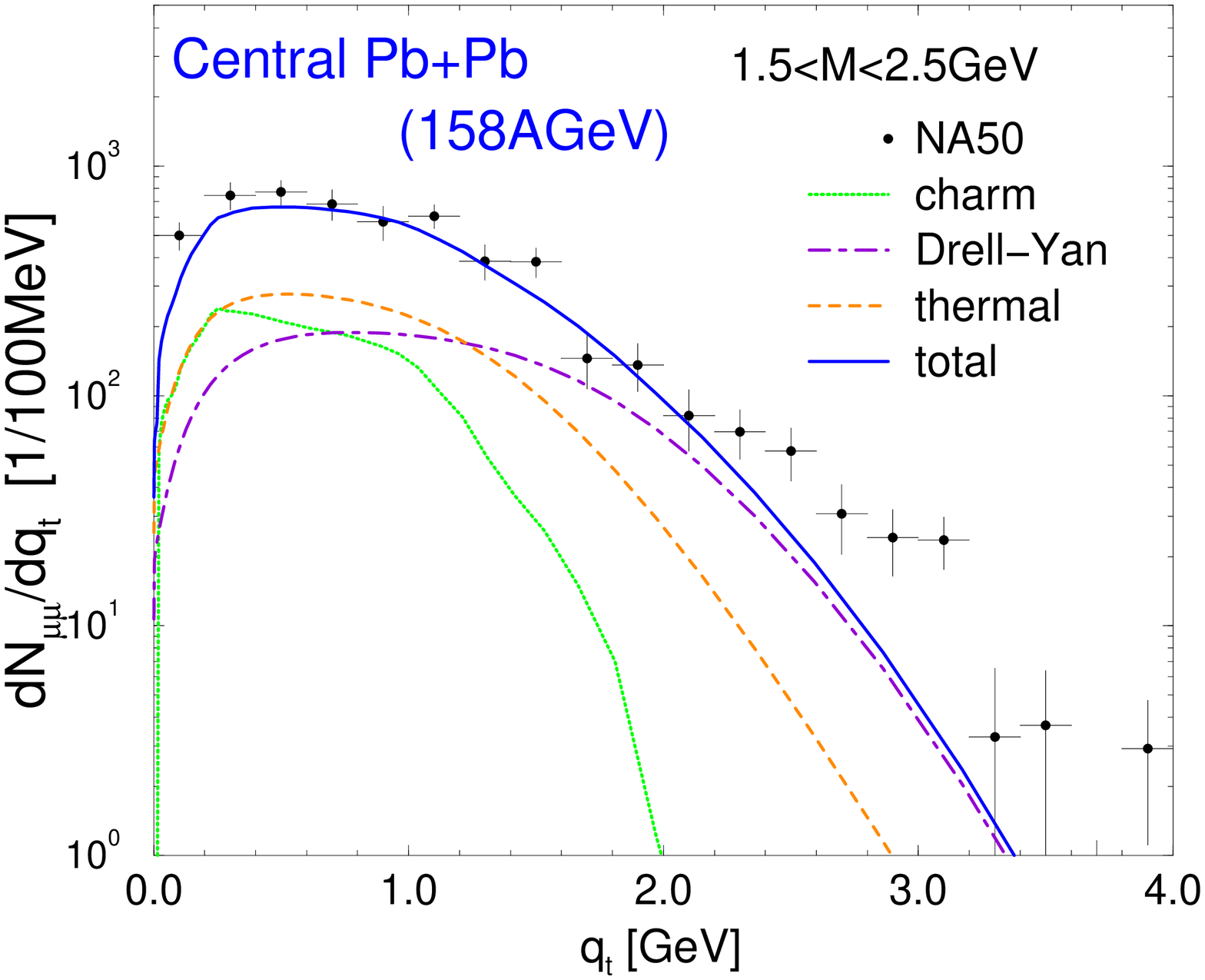,width=6.5cm,angle=-90}
\ece
\vspace{-0.7cm}
\caption{{\small Dimuon invariant-mass (left panel) and
transverse-momentum (right panel) spectra confronted with NA50 data
from central Pb(158~AGeV)+Pb collisions. In addition to the calculated
thermal and DY contributions the open-charm yield as arising from
simulations by NA50~\protect\cite{Bord99} has been included.}}
\label{fig_dlna50}
\end{figure}
The thermal dilepton spectra are then computed as
\beq
\frac{dN_{\mu\mu}^{therm}}{dM}= \int\limits_0^{t_{fo}} dt \ V_{FB}(t) 
\int \frac{M d^3q}{q_0} \ \frac{d^8N_{\mu\mu}}{d^4xd^4q}(M,q;T) \  
\left[e^{\mu_\pi/T}\right]^4 \ {\rm Acc}(M,q_t,y) \ 
\eeq 
including the experimental acceptance as determined above. 
Note the explicit appearance of the pion-fugacity factor to the
4-th power to appropriately account for off-equilibrium effects 
in 4-pion-type annihilation processes which dominate in the IMR 
(see right panel of Fig.~\ref{fig_helios3}).  
The final results of our calculation are displayed in 
Fig.~\ref{fig_dlna50}: the experimentally observed excess 
is reasonably well reproduced by thermal radiation 
in both invariant-mass and transverse-momentum projections (very similar
conclusions have been reached in ref.~\cite{GKP99}). The 
contribution from the QGP part of the evolution constitutes a rather 
moderate fraction of $\sim$~20\%.

%%%%%%%%%%%%%%%%%%%%%%%%%%%%%%%
\section{I-M Dileptons at RHIC}
\label{sec_rhic}
%%%%%%%%%%%%%%%%%%%%%%%%%%%%%%%
The same approach as described in the preceding section is now 
applied to central Au+Au collisions at $\sqrt{s}$=200~AGeV. 
For definiteness the charged particle multiplicity at midrapidity
has been fixed at $\langle N_{ch}\rangle$=800 with  an entropy per baryon of 
$s/n_B$=260. 
\begin{figure}[!htb]
\vspace{-1.3cm}
\bce
\epsfig{file=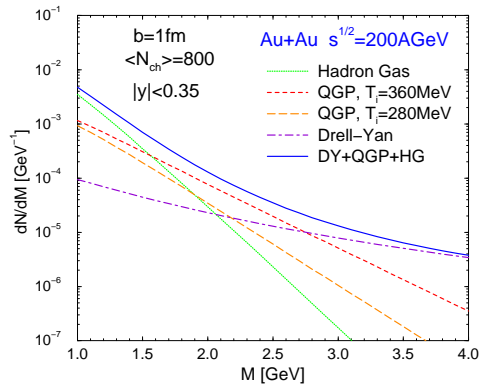,width=7cm,angle=-90}
\ece
\vspace{-0.6cm}
\caption{{\small Dilepton spectra in central Au+Au at RHIC
from equilibrated hadronic and  quark-gluon matter as well as DY 
annihilation.}}
\label{fig_rhiceq}
\end{figure}
The resulting IMR dilepton spectra are summarized 
in Fig.~\ref{fig_rhiceq}: up to $M$$\simeq$~1.5~GeV the hadron gas 
radiation dominates; in contrast to SpS conditions, the QGP contribution
dominates around 2~GeV before  DY annihilation takes over. 
Not shown is the yield from open-charm decays, which in fact 
could completely outshine the spectrum by a factor of 10 or 
so~\cite{Ga96}. If, however, charm quarks undergo appreciable
energy loss ($dE/dx$$\simeq$$-$1-2~GeV/fm) 
when propagating through hot/dense matter they might 
thermalize entailing a suppression of their contribution
above $M$=1.5~GeV by factors of $\sim$100~\cite{Shu97}. 

\begin{figure}[!htb]
\vspace{-1.2cm}
\bce
\epsfig{file=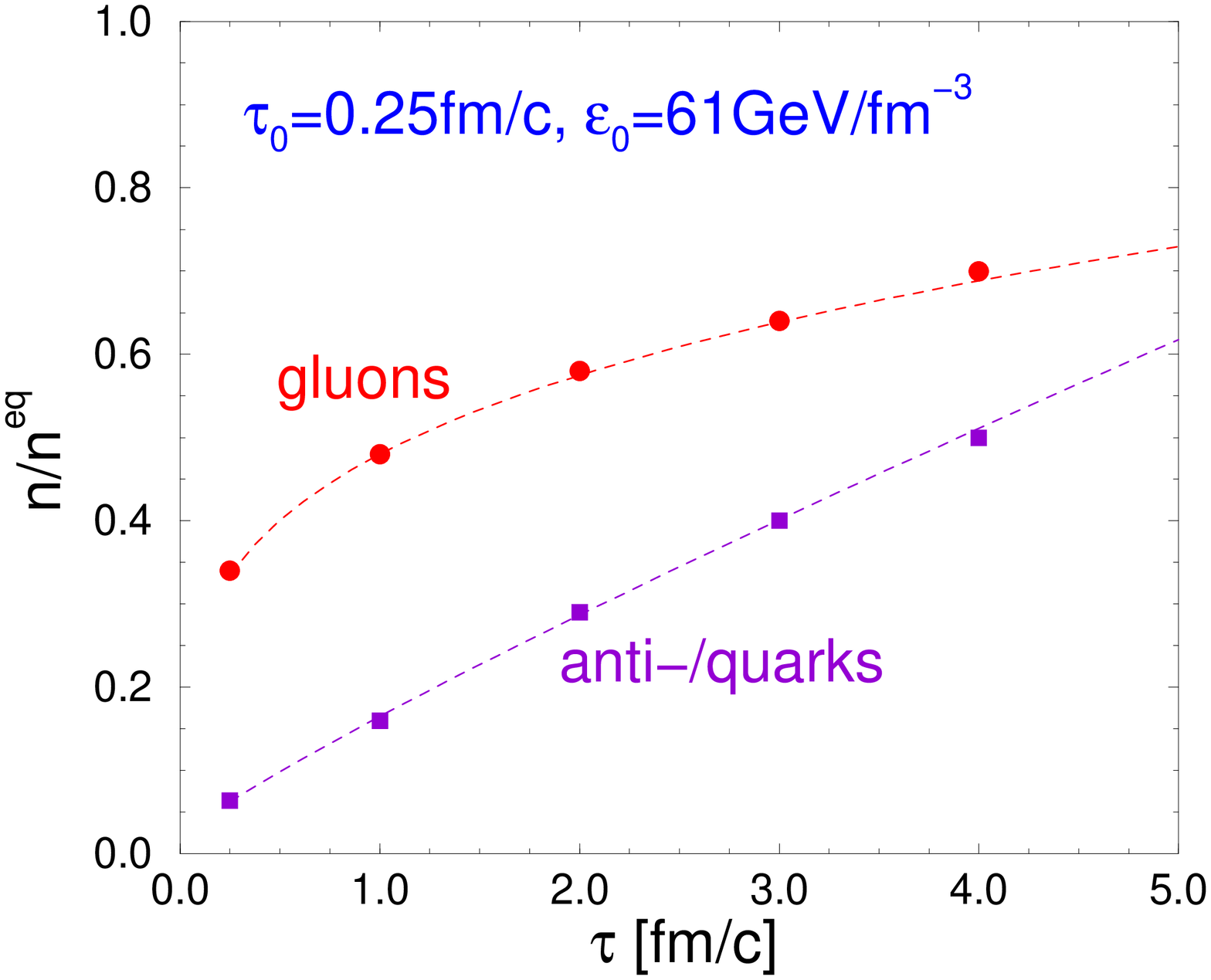,width=6.8cm,angle=-90}
\epsfig{file=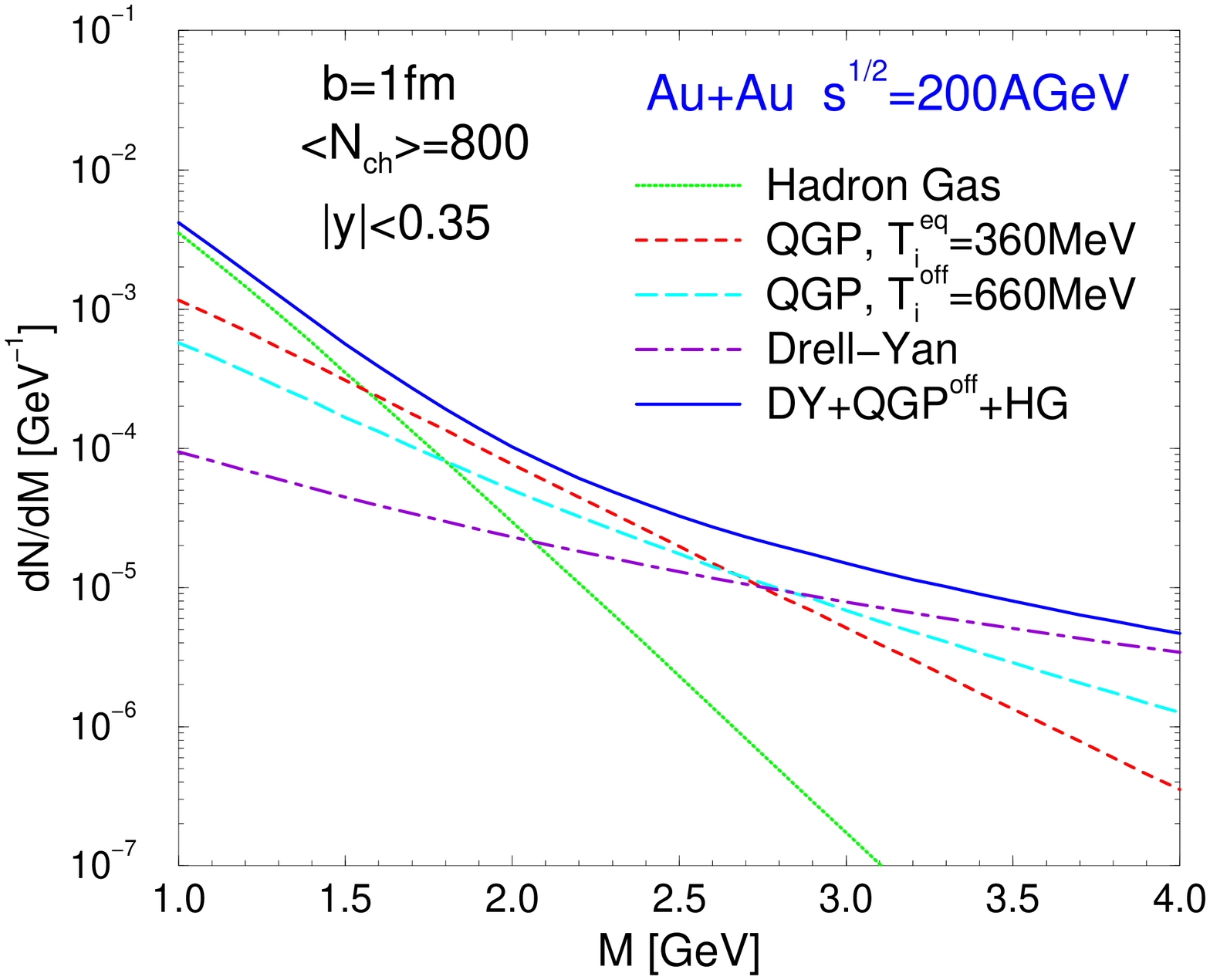,width=6.5cm,angle=-90}
\ece
\vspace{-0.7cm}
\caption{{\small Left panel: evolution of parton fugacities 
in the QGP phase at RHIC according to ref.~\protect\cite{ER99}; 
right panel: resulting dilepton spectrum (long-dashed line) 
compared to equilibrium-QGP (at equal initial entropy density), 
HG and DY yields.}}
\label{fig_rhicoff}
\vspace{-0.3cm}
\end{figure}

Another complication at RHIC energies concerns the chemical 
under-satu\-ration of gluon and especially quark densities in 
the early stages as predicted in various parton-based 
models~\cite{parton}: albeit thermalized, the parton 
distributions are characterized by fugacities  
$\lambda_i$=$n_i(T)/n_i^{eq}(T)$$<$1 ($i$=$q$,$\bar q$,$g$). 
Naively one would expect a  substantial reduction of
dilepton production in the
$q\bar q$ channel as the rate is proportional
to $\lambda_q \lambda_{\bar{q}}$. On the other hand, at
given entropy (or energy) density, an under-saturated QGP 
has a larger temperature than in chemical equilibrium
which in turn enhances the thermal emission. 
Using a parameterization of recent hydrodynamic evolution results~\cite{ER99}
(see also ref.~\cite{Ka95}) we have recalculated the plasma 
contribution starting at the same initial entropy density as in the
equilibrium scenario. The magnitude of the pertinent QGP signal 
in the final spectrum turns out to be quite similar with a somewhat
harder slope for the off-equilibrium calculation (see left panel of
Fig.~\ref{fig_rhicoff}), \ie, the reduction in the fugacities is 
largely compensated by the increase in initial temperature.

%%%%%%%%%%%%%%%%%%%%%%%%%%%%%%%%%%%
\section{Conclusions and Outlook}
\label{sec_concl}
%%%%%%%%%%%%%%%%%%%%%%%%%%%%%%%%%%%
Based on a thermal fireball model coupled with  
'standard' dilepton production rates we have shown that 
the excess observed in central Pb(158AGeV)+Pb by NA50
in the IMR can be explained by thermal radiation.
The contribution from early phases indicative for a 
QGP is moderate; however, our results 
corroborate the present  understanding of the conditions
probed at the CERN-SpS being consistent with low-mass
dilepton spectra, chemical freezeout analyses, etc.,
indicating that one is indeed producing QCD matter
in the vicinity of the expected HG-QGP phase boundary. 

The extrapolation of this approach to RHIC 
suggests that the plasma radiation exceeds 
HG- and DY-sources around $M\simeq$~2~GeV. 
Chemical off-equilibrium effects do not seem to alter this
conclusion as long as comparable initial energy densities
are reached. 
A big question mark is attached to the open-charm 
contribution, \ie, whether energy loss effects
significantly redistribute the associated dilepton yields.
Experimental input on these issues 
is eagerly awaited. \\ 

\noindent
{\bf Acknowledgment}\\
I thank E. Shuryak and B. K\"ampfer for productive discussion.
This work is supported in part by the A.-v.-Humboldt foundation
(Feodor-Lynen program) and the US-DOE under grant no. DE-FG02-88ER40388.

%%%%%%%%%%%%%%%%%%%%%%%%%%%%

\end{document}